\documentclass[aps,prl,floatfix,twocolumn,reprint,amsmath,amssymb,superscriptaddress]{revtex4-1}
\usepackage{amsfonts}
\usepackage{mathrsfs}
\usepackage{amsmath}
\usepackage{color}
\usepackage{natbib}
\usepackage{graphicx}
\usepackage{bm}
\usepackage{amssymb}
\usepackage{xspace}
\usepackage{epstopdf}
\usepackage{dcolumn}
\usepackage{longtable}
\usepackage{array}
\usepackage{makecell}
\usepackage{tabulary}
\usepackage{multirow}
\newcolumntype{C}[1]{>{\centering\arraybackslash}p{#1}}
\newcolumntype{L}[1]{>{\flushleft\arraybackslash}p{#1}}

\usepackage{multirow}
\usepackage{epsfig}
\usepackage[normalem]{ulem}
\usepackage{amsmath}
\usepackage{braket}
\usepackage{bbold}
\usepackage{natbib}

\usepackage[colorlinks=true, letterpaper=true, pdfstartview=FitV, linkcolor=blue, citecolor=blue, urlcolor=blue]{hyperref}

\makeatletter

\newcommand{\Rmnum}[1]{\expandafter\@slowromancap\romannumeral #1@}
\makeatother

\begin{document}
	
	\title{Disorder- and Topology-Enhanced Fully Spin-Polarized Currents in Nodal Chain Spin-Gapless Semimetals}
	
	\author{Xiaodong Zhou}
	\affiliation{Centre for Quantum Physics, Key Laboratory of Advanced Optoelectronic Quantum Architecture and Measurement (MOE),School of Physics, Beijing Institute of Technology, Beijing, 100081, China}
	\affiliation{Beijing Key Lab of Nanophotonics $\&$ Ultrafine Optoelectronic Systems, School of Physics, Beijing Institute of Technology, Beijing, 100081, China}    
	
	\author{Run-Wu Zhang}
	\affiliation{Centre for Quantum Physics, Key Laboratory of Advanced Optoelectronic Quantum Architecture and Measurement (MOE),School of Physics, Beijing Institute of Technology, Beijing, 100081, China}
	\affiliation{Beijing Key Lab of Nanophotonics $\&$ Ultrafine Optoelectronic Systems, School of Physics, Beijing Institute of Technology, Beijing, 100081, China}    
	
	\author{Xiuxian Yang}
	\affiliation{Centre for Quantum Physics, Key Laboratory of Advanced Optoelectronic Quantum Architecture and Measurement (MOE),School of Physics, Beijing Institute of Technology, Beijing, 100081, China}
	\affiliation{Beijing Key Lab of Nanophotonics $\&$ Ultrafine Optoelectronic Systems, School of Physics, Beijing Institute of Technology, Beijing, 100081, China}
	
	\author{Xiao-Ping Li}
	\affiliation{Centre for Quantum Physics, Key Laboratory of Advanced Optoelectronic Quantum Architecture and Measurement (MOE),School of Physics, Beijing Institute of Technology, Beijing, 100081, China}
	\affiliation{Beijing Key Lab of Nanophotonics $\&$ Ultrafine Optoelectronic Systems, School of Physics, Beijing Institute of Technology, Beijing, 100081, China}   
	
	\author{Wanxiang Feng}
	\email{wxfeng@bit.edu.cn}
	\affiliation{Centre for Quantum Physics, Key Laboratory of Advanced Optoelectronic Quantum Architecture and Measurement (MOE),School of Physics, Beijing Institute of Technology, Beijing, 100081, China}
	\affiliation{Beijing Key Lab of Nanophotonics $\&$ Ultrafine Optoelectronic Systems, School of Physics, Beijing Institute of Technology, Beijing, 100081, China}
	
	\author{Yuriy Mokrousov}
	\email{y.mokrousov@fz-juelich.de}
	\affiliation{Peter Gr\"unberg Institut and Institute for Advanced Simulation, Forschungszentrum J\"ulich and JARA, 52425 J\"ulich, Germany}
	\affiliation{Institute of Physics, Johannes Gutenberg University Mainz, 55099 Mainz, Germany}
	
	\author{Yugui Yao}
	\email{ygyao@bit.edu.cn}
	\affiliation{Centre for Quantum Physics, Key Laboratory of Advanced Optoelectronic Quantum Architecture and Measurement (MOE),School of Physics, Beijing Institute of Technology, Beijing, 100081, China}
	\affiliation{Beijing Key Lab of Nanophotonics $\&$ Ultrafine Optoelectronic Systems, School of Physics, Beijing Institute of Technology, Beijing, 100081, China}  
	
	\date{\today}
	
	\begin{abstract}
		Recently discovered high-quality nodal chain spin-gapless semimetals $M$F$_3$ ($M$ = Pd, Mn)  feature an ultra-clean nodal chain in the spin up channel residing right at the Fermi level and displaying a large spin gap leading to a 100\% spin-polarization of transport properties. Here, we investigate both intrinsic and extrinsic contributions to anomalous and spin transport in this class of materials. The dominant intrinsic origin is found to originate entirely from the gapped nodal chains without the entanglement of any other trivial bands. The side-jump mechanism is predicted to be negligibly small, but intrinsic skew-scattering enhances the intrinsic Hall and Nernst signals significantly, leading to large values of respective conductivities. Our findings open a new material platform for exploring strong anomalous and spin transport properties in magnetic topological semimetals.
	\end{abstract}
	\maketitle
	
	
	\textit{\textcolor{black}{Introduction.---}}
	After the discovery of magnetic topological insulators~\cite{R-Yu2010,CZ-Chang2013}, different types of magnetic topological states ranging from insulators to semimetals have emerged~\cite{XG-Wan2011,G-Xu2011,C-Fang2014,ZJ-Wang2016,XL-Wang2016,PZ-Tang2016,Smejkal2017,Kuroda2017,QN-Xu2018,DF-Liu2019,Morali2019,Belopolski2019,BJ-Feng2019,SM-Nie2020,RW-Zhang2020,YF-Xu2020,JH-Li2019,Otrokov20191,Otrokov20192,DQ-Zhang2019,Y-Gong2019,YJ-Hao2019,H-Li2019,YJ-Chen2019,C-Liu2020,YJ-Deng2020}. This brings new vitality to the ideas evolving around the next generation of dissipationless  spintronic devices benefiting from exotic anomalous and spin transport properties. The Hall currents of charge $J_H^A$ can be  generated in magnetic materials either by an applied electric field $E$ or a thermal gradient $-\nabla T$, known correspondingly  as the anomalous Hall effect (AHE)~\cite{YG-Yao2004,Nagaosa2010} and anomalous Nernst effect (ANE)~\cite{D-Xiao2006,D-Xiao2010}.  Magnetic topological semimetals provide a prominent advantage to enhance anomalous Hall conductivity (AHC) and/or anomalous Nernst conductivity (ANC) driven by the divergent Berry curvature of gapped Weyl points or nodal lines (NLs), as reported previously for various systems theoretically and experimentally~\cite{Ikhlas2017,XK-Li2017,Sakai2018,Q-Wang2018,EK-Liu2018,Kim2018,Noky2019,Noky2020,Guin2019,Guin2019a,Sakai2020,Minami2020,PG-Li2020,HY-Yang2020,Guguchia2020,HB-Zhou2020,Yanagi2021}.  On the other hand, the Hall currents of spin $J_H^S$ $-$ i.e. the spin counterparts of $J_H^A$ also known as the spin Hall effect (SHE)~\cite{Sinova2015} and spin Nernst effect (SNE)~\cite{SG-Cheng2008} $-$ are also largely driven by
	topological nodal features which give rise to large spin Hall conductivity (SHC) and spin Nernst conductivity (SNC) in non-magnetic topological semimetals~\cite{Sahin2015,Y-Sun2016,Derunova2019,Bhowal2019,Y-Zhang2020,Y-Yen2020,Prasad2020,Taguchi2020,HJ-Xu2020,ZD-Chi2020,Ng2021,WJ-Hou2021,Leiva2021,K-Tang2021}.
	
	In the realm of emergent anomalous and spin transport in topological semimetals, however, many issues still have to be addressed.  One of the most prominent aspects is the role played by disorder-induced extrinsic mechanisms.  It is well known that both AHC $\sigma_{xy}$ and ANC $\alpha_{xy}$ can be decomposed into three different contributions (SHC $\sigma_{xy}^s$ and SNC $\alpha_{xy}^s$ are also the case)~\cite{Weischenberg2011,Weischenberg2013,Czaja2014,Tauber2012}:
	\begin{equation}
		\alpha_{xy}^{tot} (\sigma_{xy}^{tot}) = \alpha_{xy}^{int} (\sigma_{xy}^{int}) + \alpha_{xy}^{sk} (\sigma_{xy}^{sk}) + \alpha_{xy}^{sj} (\sigma_{xy}^{sj}).
	\end{equation}
	The first term is the so-called intrinsic ($int$) contribution, which can be well described by Berry phase theory~\cite{YG-Yao2004,Nagaosa2010}, and which was the focus of previous studies on magnetic topological semimetals~\cite{Ikhlas2017,XK-Li2017,Sakai2018,Q-Wang2018,EK-Liu2018,Kim2018,Noky2019,Noky2020,Guin2019,Guin2019a,Sakai2020,Minami2020,PG-Li2020,HY-Yang2020,Guguchia2020}.  The second and the last terms are the disorder-driven extrinsic contributions referred to as the skew-scattering ($sk$)~\cite{Smit1955,Smit1958} and side-jump ($sj$)~\cite{Berger1970}, respectively, and whose role in Hall effects of magnetic topological semimetals received very little attention so far, besides several very recent experimental~\cite{LC-Ding2019,JL-Shen2020,HY-Yang2020a} and model studies~\cite{Burkov2014,CZ-Chen2015,Shapourian2016,Ado2017,Keser2019,Papaj2021}.  Another challenge is to draw a clear correlation between the topological characterization and the magnitude of transport properties. Most of the previously reported materials suffer from a ``contaminated" band dispersion around the Fermi level. Moreover, the situation is complicated by the fact that often the band topology is formed by fermions of opposite spin with parabolic dispersion, which greatly decreases the current spin-polarization and the carrier Fermi velocity in real spintronic devices. 
	
	In this Letter, we address the above two issues directly.  Using first-principles calculations, we collect  all contributions to the AHE and ANE as well as the SHE and SNE in the recently proposed novel nodal chain spin-gapless semimetals (NCSGSMs) $M$F$_3$ ($M$ = Pd, Mn)~\cite{RW-Zhang2020}, which feature an ultra-clean nodal chain residing right at the Fermi level, providing an ultrahigh Fermi velocity and 100\% spin-polarization simultaneously. This provides us with a prefect platform to clearly identify pure topological contributions to Hall transport. We show that such a remarkable electronic structure inevitably gives rise to large Hall effects. We further uncover the intrinsic mechanism as the main underlying  physical origin of the large AHC and ANC, resulting from the gapped nodal chain-induced large Berry curvature. We find the side-jump to be negligibly small but discover that the intrinsic skew-scattering plays an important role for the overall signal. Our work provides a foundation for educated design of large pure spin-polarized Hall currents for future ``green" spintronics.
	
	\begin{figure}
		\includegraphics[width=\columnwidth]{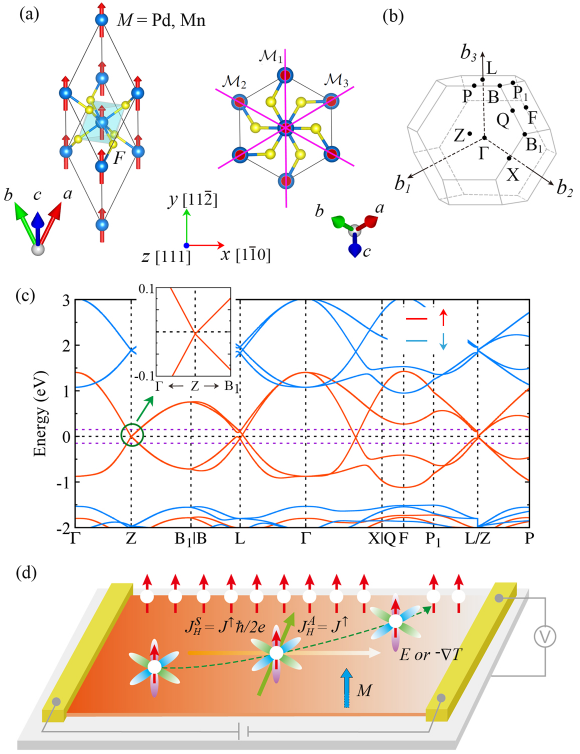}
		\caption{Nodal chain spin gapless semimetals (NCSGSMs) and their fully spin-polarized currents. (a) The crystal structure of $M$F$_3$ ($M$ = Pt, Mn), and the view of (111) plane. The blue spheres are magnetic $M$ atoms, whereas yellow spheres are nonmagnetic F atoms. The red arrows label the spin magnetization aligned along [111] direction. The pink lines denote  three mirror planes. The sketch of the  Brillouin zone is shown in (b), and (c) shows spin-polarized band structure without SOC, where the inset is a zoom into the bands near the $Z$ point. The horizontal violet dashed lines mark the considered energy range of anomalous and spin transport. (d) Schematic illustration of fully spin-polarized Hall current induced by AHE (SHE) and ANE (SNE) in a NCSGSM (indicated with a hexagonal petal), when an electric field or a temperature gradient field is applied along the longitudinal direction.}
		\label{crystal}
	\end{figure}
	
	\begin{figure*}
		\includegraphics[width=0.9\textwidth]{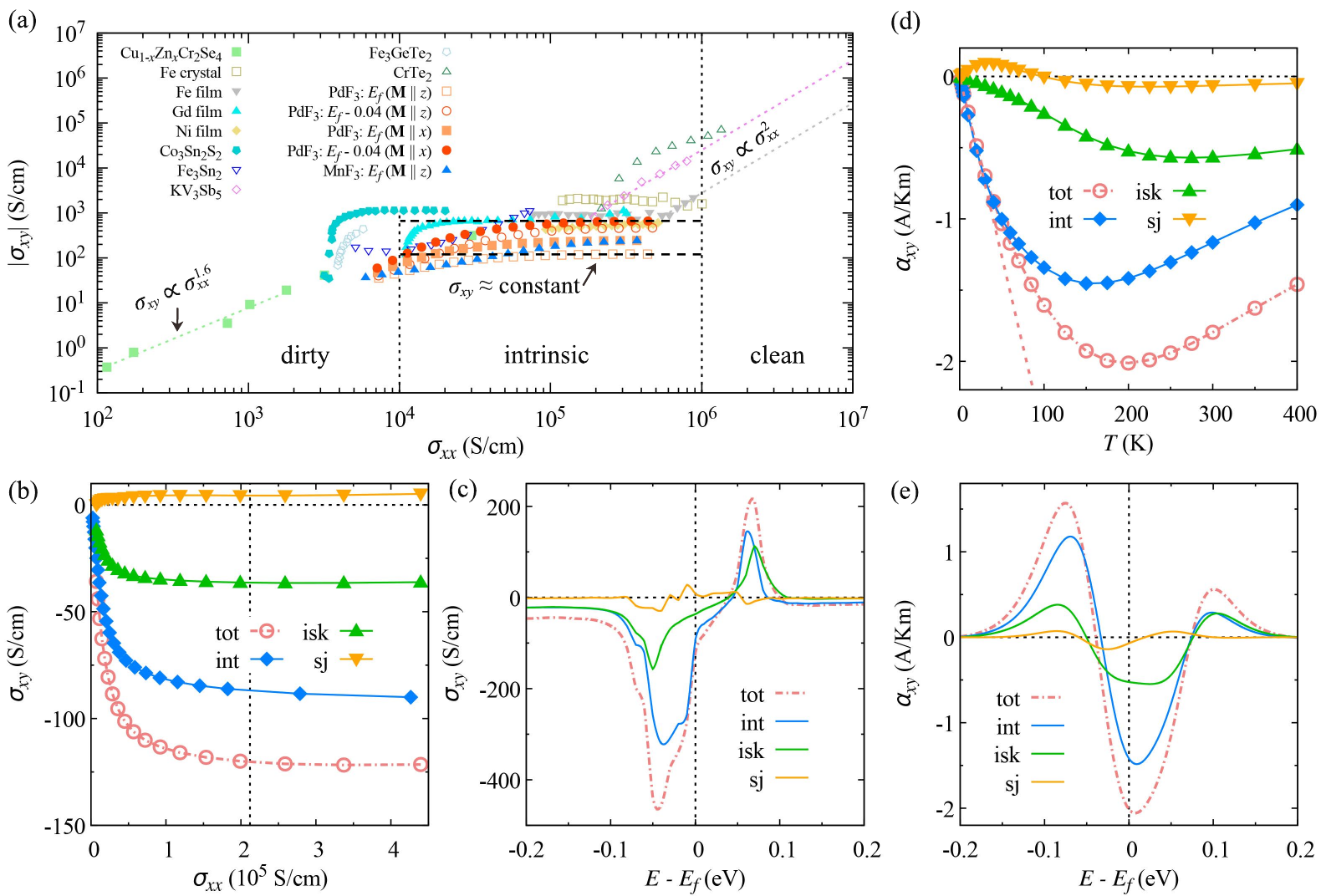}
		\caption{Large spin-pure AHC and ANC. (a) AHC $\sigma_{xy}$ versus $\sigma_{xx}$ for  PdF$_3$ at the true Fermi energy $E_f$ and at $E_f - 0.04$\,eV, with the magnetization $\textbf{M}$  along the $z$ or $x$ direction, ranging across intrinsic from dirty to clean regimes. Note that when $\textbf{M} \parallel x$, the nonzero component of AHC is $\sigma_{yz}$.  The data of MnF$_3$ for $\textbf{M}\parallel z$ is also plotted.  In the plot, data for various ferromagnets are shown for comparison, as reported in previous works~\cite{Miyasato2007,SY-Yang2020,M-Huang2021}. (b-c) Disorder-related contributions to the AHC ($\sigma_{xy}^{int}$, $\sigma_{xy}^{isk}$, $\sigma_{xy}^{sj}$, and the total $\sigma_{xy}$) as a function of  $\sigma_{xx}$ (at $E_f$) and energy, respectively.  In (c), the intrinsic AHC is calculated at the clean limit, while the extrinsic AHC is evaluated by incorporating a Gaussian disorder potential with a weak disorder parameter (1.83 eV$^2$bohr$^3$)~\cite{SuppMater}, corresponding to $\sigma_{xx} = 2.12\times 10^5$ S/cm, at where the extrinsic $\sigma_{xy}$ reaches to a plateau, indicated by a vertical dashed line in (b).  (d-e) Disorder-related contributions to the ANC ($\alpha_{xy}^{int}$, $\alpha_{xy}^{isk}$, $\alpha_{xy}^{sj}$, and the total $\alpha_{xy}$) for various temperatures $T$ and energy, computed at $E_f$ and for $T=200$\,K, respectively.}
		\label{AHE_ANE}
	\end{figure*}
	
\begin{figure*}
		\includegraphics[width=0.9\textwidth]{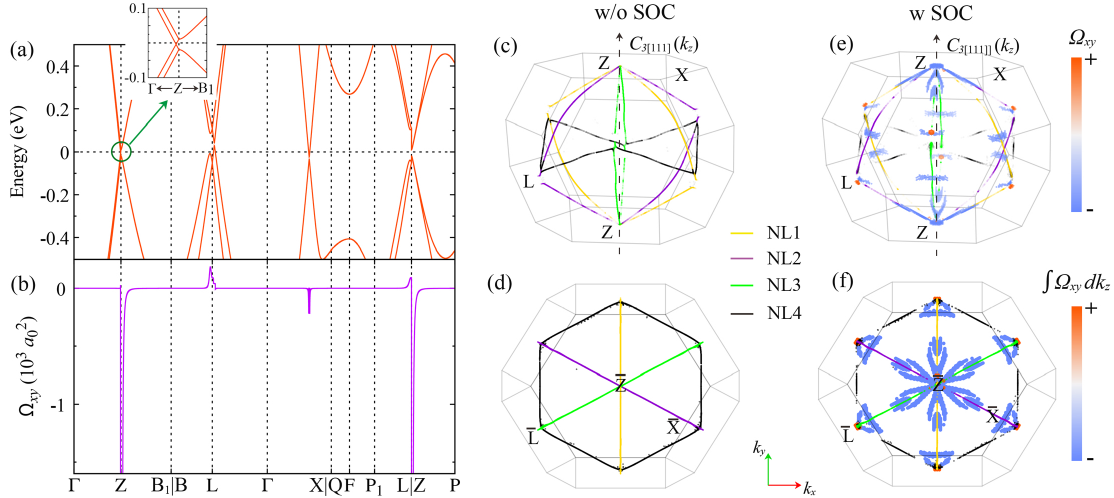}
		\caption{The underlying physical origin of large AHC and ANC. (a) The band structure of PdF$_3$ with SOC, with the inset zooming into the bands near the $Z$ point. (b) Berry curvature $\Omega_{xy}$ along high symmetry lines (singular value at $Z$ point reaches 0.46$\times 10^5$ bohr$^2$). (c) The distribution of nodal lines in the Brillouin zone, and (d) projected on $k_x$-$k_y$ plane. (e) $\Omega_{xy}$ and nodal lines in the Brillouin zone, and (f) projected to the $k_x$-$k_y$ plane where $\Omega_{xy}$ is integrated along the $k_z$ direction, i.e., $\int \Omega_{xy}(k_x, k_y, k_z)dk_z$.  To present main features in the distribution of $\Omega_{xy}$, the values of less than 200 bohr$^2$ are not shown.}
		\label{berrycurvature}
	\end{figure*}

	\textit{\textcolor{black}{High-quality candidate hosting large fully spin-polarized current.---}} A crucial issue for addressing topological contributions to anomalous transport is screening out the influence of trivial bands. To tackle this, we focus on a promising candidate platform $-$ the recently proposed spin gapless semiconductors or semimetals~\cite{XL-Wang2008,RW-Zhang2020,YL-Jiao2017,XL-Wang2016,Q-Gao2019}.  Among these, we select an outstanding example, the rhombohedral transition metal trifluorides $M$F$_3$ ($M$ = Pd, Mn)~\cite{YL-Jiao2017,RW-Zhang2020} (Fig.~\ref{crystal}(a)).  The ferromagnetic ground state is confirmed by performing a series of spin spiral calculations, as shown in Figs.~\textcolor{blue}{S1} and~\textcolor{blue}{S2} in Supplemental Material~\cite{SuppMater}.  The high Curie temperature ($\geq$ 450 K) is estimated from  mean field approximation, Monte Carlo simulations, and Landau-Lifshitz-Gilbert spin dynamics (Fig.~\textcolor{blue}{S3}~\cite{SuppMater}).  $M$F$_3$ display a linear semimetallic band structure in the spin up channel, while exhibiting a large indirect band gap (2.46 eV for PdF$_3$,  6.43 eV for MnF$_3$) in the spin down channel (Fig.~\ref{crystal}(c) and Fig.~\textcolor{blue}{S4}~\cite{SuppMater}), thus enjoying a 100\% spin-polarization of the states at the Fermi energy.  The spin-up electronic structure around the Fermi level is formed by two types of cross-connection modes. The one mode comprises three accidentally formed NLs (NL$_1$, NL$_2$, NL$_3$) (see Fig.~\ref{berrycurvature}(c-d)), which are positioned in three mirror planes ($\mathcal{M}_1, \mathcal{M}_2$, $\mathcal{M}_3$) (see Fig.~\ref{crystal}(a)) and are pinned at the two $Z$  points [(0.5,0.5,0.5), ($-$0.5,$-$0.5,$-$0.5)].  The other  mode, a ``snakelike" structure with six corners at the $L$ points (NL$_4$), crosses the former three NLs transversely.
	
	The ultra-clean topological nodal lines (NL$_{1-4}$) not only perfectly avoid the entanglement of trivial bands, but can generate the much desired fully spin-polarized Hall current based on AHE and ANE as well as their spin counterparts SHE and SNE (Fig.~\ref{crystal}(d)). Accordingly, the anomalous and spin Hall currents can be written down as follows~\cite{Tauber2012}:
	\begin{eqnarray}
		J_H^A &=& J^\uparrow + J^\downarrow = J^\uparrow,  \label{eq:J_H^A} \\
		J_H^S &=& (J^\uparrow - J^\downarrow)\dfrac{\hbar}{2e} =\dfrac{\hbar}{2e} J^\uparrow \label{eq:J_H^S},
	\end{eqnarray}
	since the spin-down bands reside far away from the Fermi level and do not contribute to the Hall effect. Respectively, the SHC and SNC are given by $\sigma_{xy}^s = \frac{\hbar}{2e}\sigma_{xy}$ and $\alpha_{xy}^s = \frac{\hbar}{2e}\alpha_{xy}$, as also confirmed by first-principles calculations (see Fig.~\textcolor{blue}{S5}~\cite{SuppMater}).
	
	Next we proceed to explore these physical phenomena quantitatively. First, to confirm the leading mechanism of AHC in $M$F$_3$, the computed variation of $\sigma_{xy}$ (the superscript $tot$ is omitted in the following discussion) with $\sigma_{xx}$ at $E = E_f$ and $E = E_f-0.04$ eV is plotted in Fig.~\ref{AHE_ANE}(a) for the magnetization $\textbf{M}$ being along the $z$-axis ([111] direction) and $x$-axis ([1$\bar{1}$0] direction), respectively.  To do this, we use the implementation of uncorrelated disorder scattering formalism from the first-principles (see Eqs. \textcolor{blue}{(S5)-(S8)}~\cite{SuppMater}). By analysing the dependency of $\sigma_{xy}$ on $\sigma_{xx}$, different scaling relations have been proposed for a variety of ferromagnets~\cite{Miyasato2007,DZ-Hou2015,SY-Yang2020,Park2020,M-Huang2021}: $\sigma_{xy} \propto \sigma_{xx}^{1.6}$ in the dirty regime ($\sigma_{xx} < 10^4$ S/cm), nearly constant in the intrinsic regime ($\sigma_{xx} \sim10^4-10^6$ S/cm), and $\sigma_{xy} \propto \sigma_{xx}^{2}$ or $\sigma_{xx}^{1}$ in the clean regime ($\sigma_{xx} > 10^6$ S/cm).  From Fig.~\ref{AHE_ANE}(a), one can see that $M$F$_3$ is located within the intrinsic regime, and $\sigma_{xy}$ exhibits a nearly constant plateau for both $E = E_f$ and $E = E_f-0.04$ eV for $10^5$  $<\sigma_{xx} < 10^6$ S/cm, in accordance with the above scaling relation. In other words, the AHC is dominated by the intrinsic mechanism. To reveal this observation more clearly, the component-resolved AHC of PdF$_3$ with $E = E_f$ and $E = E_f-0.04$ eV when $\textbf{M} \Vert z$ are shown in Figs.~\ref{AHE_ANE}(b) and~\textcolor{blue}{S6} as $\sigma_{xx}$ is varied. The intrinsic part dominates the shape of AHC, while  the intrinsic skew-scattering also plays an important role, contributing by about one-third of $\sigma_{xy}$. In contrast, we find the side-jump to be negligibly small. Note that although the density of states for semimetals is usually small at the Fermi level, the longitudinal conductivity $\sigma_{xx}$ can reach up to $10^3\sim10^4$ S/cm for WTe$_2$~\cite{SY-Shi2019,B-Zhao2020,L-Wang2016} and Co$_3$Sn$_2$S$_2$~\cite{EK-Liu2018,Q-Wang2018}, $10^4 \sim 10^7$ S/cm for ZrSiS~\cite{Ali2016}, and $10^4 \sim10^6$ S/cm for our cases $M$F$_3$.
	
	The energy evolution of $\sigma_{xy}$,  shown in Fig.~\ref{AHE_ANE}(c), reveals considerable values in the range of [$-$0.1, 0.1] eV, near the band crossing points, which can be easily accessed by current experimental techniques such as angle-resolved photoemission spectroscopy~\cite{DF-Liu2019}. Another prominent feature of Fig.~\ref{AHE_ANE}(c) is a large variation of $\sigma_{xy}$ with  energy. According to the low-temperature Mott relation, which relates the ANC to the energy derivative of the AHC~\cite{D-Xiao2006},
	\begin{equation}\label{eq:Mott2}
		\alpha_{xy} = -\dfrac{\pi^{2}k_{B}^{2}T}{3e}\left.\frac{d\sigma_{xy}}{dE}\right|_{E=E_f}
	\end{equation}
	where $k_B$, $T$, and $e$ are Boltzmann constant, temperature, and elementary charge, respectively,  one can expect a large ANC near the Fermi energy. Using the generalized Mott formula~\cite{SuppMater}, we compute the ANC and show the component-resolved data in Figs.~\ref{AHE_ANE}(d) and~\ref{AHE_ANE}(e). The temperature dependence of ANC shows that the low-temperature Mott relation is valid up to about 40 K, and one can indeed observe  an ANC which is much larger than that in traditional ferromagnets (typically $|\alpha_{xy}| = 0.01-1$\,A/Km~\cite{Sakai2018}) once $T\geqslant$ 50 K. Remarkably, a large $\alpha_{xy}$ of about 2 A/Km is observed at $T$ = 200 K, which is by far larger than that in conventional ferromagnets. Another prominent advantage of PdF$_3$ is that the peak of $\alpha_{xy}$ is positioned right at the true $E_f$, as seen in Fig.~\ref{AHE_ANE}(e), while the range of energies with large $\alpha_{xy}$  is wide enough for large ANE to be easily detected experimentally. 
	
	\begin{figure}
		\includegraphics[width=\columnwidth]{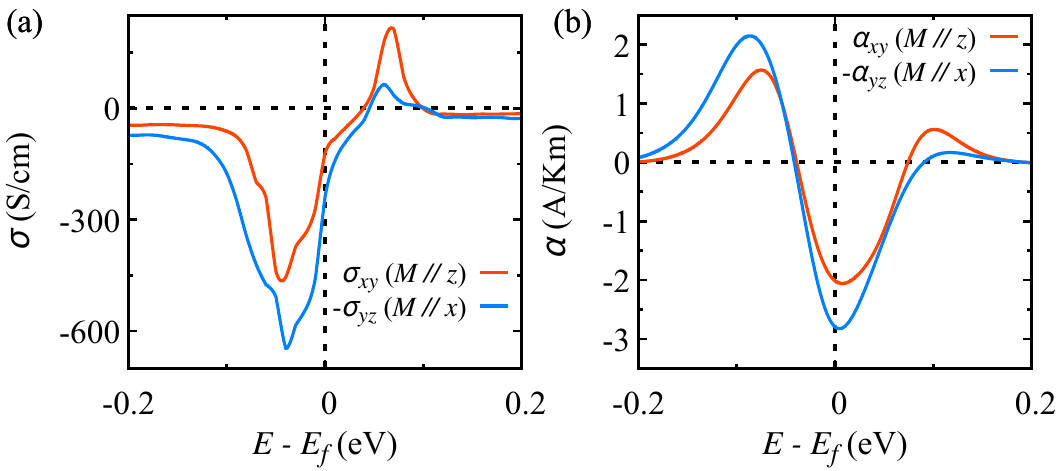}
		\caption{Anisotropic AHC and ANC. (a-b) Total AHC and ANC ($T$ = 200 K) for PdF$_3$ when the magnetization $\textbf{M}$ is aligned along the $z$- and $x$-axes.}
		\label{Anisotropic ANE}
	\end{figure}
	
	\textit{\textcolor{black}{The underlying physical origin of strong transport properties.---}} Next, we  uncover the underlying physical origin of the large AHC and ANC in PdF$_3$ (similar analysis can be performed for SHC and SNC). The band structure of PdF$_3$ with SOC together with the Berry curvature $\Omega_{xy}$ along high symmetry lines is shown in Figs.~\ref{berrycurvature}(a-b). Clearly, the slightly gapped crossing points generate large $\Omega_{xy}$ with negligible contributions at other ``trivial" regions. Specifically, a pronounced negative peak is found near the $Z (0.5, 0.5, 0.5)$ point, which is the rotation-invariant point of three glide mirrors ($\mathcal{\widetilde{M}}_1$, $\mathcal{\widetilde{M}}_2$, and $\mathcal{\widetilde{M}}_3$ $-$ the combined symmetries of three mirrors, Fig.~\ref{crystal}(a), with translational symmetry), and which hosts  a fourfold degeneracy in the absence of SOC, ensured by three glide mirrors and $C_{3[111]}$ symmetry,  Fig.~\ref{crystal}(c)~\cite{RW-Zhang2020}. The SOC breaks  three glide mirrors, and the fourfold degeneracy is split into a gapped group of states at $Z$ point, and an accidental degeneracy along the $\Gamma Z$ direction. The former rather than the latter is responsible for the large $\Omega_{xy}$. Indeed, the distribution of $\Omega_{xy}$ in the Brillouin zone (Fig.~\ref{berrycurvature}(e)) indicates that the hot spots are mainly distributed near the gapped nodal lines. Further, we integrate  $\Omega_{xy}$ along the $k_z$ direction (Fig.~\ref{berrycurvature}(f)), which shows prominent features near the $\bar{Z}, \bar{L}$ points and along the $\bar{Z}\bar{X}$ direction. We stress that large AHE and ANE predicted here differ from previous studies~\cite{Sakai2018,Sakai2020}, which require a simultaneous enhancement of the Berry curvature and density of states created by a large Fermi surface with Weyl points or a flat nodal line.  In contrast, for NCSGSMs, the density of states is nearly vanishing at the Fermi level.  Hence the large AHC and ANC predicted here are driven by pure topological characteristics.
	
	To further confirm the topological origin of transport in PdF$_3$, we also consider the case of the magnetization directed along other directions,~e.g.,~$x$- and $y$-axes (see Figs.~\ref{Anisotropic ANE},~\textcolor{blue}{S7--S9}~\cite{SuppMater}). The results show that the spectral features of the symmetry-allowed AHC and ANC are nearly the same as those for the case of magnetization oriented along the $z$-axis, while there is a large difference in magnitude, which indicates a strongly anisotropic anomalous and spin transverse transport in the $zx$ or $zy$ plane. Remarkably, a large AHC of 646 S/cm and ANC of 2.8 A/Km are found (Fig.~\ref{Anisotropic ANE}), among which the latter is particularly striking owing to the magnitude larger than that in the famous kagome magnet Co$_3$Sn$_2$S$_2$~\cite{Guin2019a} while approaching the largest recorded experimental values of about 4 A/Km in Co$_2$MnGa~\cite{Sakai2018} and 5.2 A/Km in Fe$_3$Ga~\cite{Sakai2020}. In Figs.~\textcolor{blue}{S7} and~\textcolor{blue}{S9}, we present evidence that this enhancement originates from the intrinsic contribution, mediated by the reduction of magnetic symmetries from $R\bar{3}c^{\prime}$ for magnetization along $z$-axis to $C2/c$ for $x$-axis ($C2^{\prime}/c^{\prime}$ for $y$-axis), that  splits  the nodal lines further and gives rise to Berry curvature's amplification. In contrast, the change in extrinsic contributions is negligible. This further supports the topological origin of large anomalous and spin transport in PdF$_3$.
	
	Finally, we  also investigated transverse transport in other NCSGSMs, in particular in MnF$_3$~\cite{YL-Jiao2017,RW-Zhang2020}. Large AHC (SHC) and ANC (SNC) are also found due to the similar topological band structure and Berry curvature distribution (see Figs.~\textcolor{blue}{S4} and~\textcolor{blue}{S5}~\cite{SuppMater}).  We believe that large fully spin-polarized currents can be also found in a large family of spin-gapless semimetals or semiconductors~\cite{XL-Wang2008,RW-Zhang2020,YL-Jiao2017,XL-Wang2016,Q-Gao2019}.
	
	The authors thank S. Di Napoli, Fawei Zheng, Jin-Jian Zhou, Chaoxi Cui, Shifeng Qian, and Jin Cao for fruitful discussions. This work is supported by the National Key R\&D Program of China (Grant No. 2020YFA0308800), the National Natural Science Foundation of China (Grant Nos. 11734003, 11874085, and 12047512), the Science \& Technology Innovation Program of Beijing Institute of Technology (Grant No. 2021CX01020), and the Project Funded by China Postdoctoral Science Foundation (Grant Nos. 2020M680011 and 2021T140057).  Y.M. acknowledges the Deutsche Forschungsgemeinschaft (DFG, German Research Foundation) - TRR 288 - 422213477 (project B06). Y.M., W.F., and Y.Y. acknowledge the funding under the Joint Sino-German Research Projects (Chinese Grant No. 12061131002 and German Grant No. 1731/10-1) and the Sino-German Mobility Programme (Grant No. M-0142).
	

%

\end{document}